\RequirePackage[l2tabu, orthodox]{nag}
\documentclass[12pt, notitlepage]{article}
\usepackage{graphicx}
\usepackage{amssymb,amsmath,amsfonts,palatino,amsthm}
\usepackage{bbold}
\usepackage{booktabs}
\usepackage{braket}
\usepackage{cite}
\usepackage[left=2cm,right=2cm,top=2cm,bottom=2cm]{geometry}
\usepackage[utf8]{inputenc}
\usepackage{microtype}
\usepackage{ragged2e}
\usepackage{slashed}
\usepackage[usenames,dvipsnames]{xcolor}
\newtheorem{theorem}{Theorem}

\newcommand{\f}{\begin{equation}}
\newcommand{\ff}{\end{equation}}
\usepackage{fancyhdr}
\usepackage[numbers, square, sort, comma]{natbib}
\usepackage[colorlinks=true, pdfborder={0 0 0}, allcolors = black, bookmarks = true, bookmarksnumbered = true, linktoc = 'section']{hyperref}

\usepackage{cleveref}
\usepackage{subfigure}

\title{\textbf{Fermion Doubling in Loop Quantum Gravity}  \\}
\author{Jacob Barnett\thanks{jbarnett@perimeterinstitute.ca}\, and Lee Smolin\thanks{lsmolin@perimeterinstitute.ca} \\ \\
Perimeter Institute for Theoretical Physics,\\
31 Caroline Street North, Waterloo, Ontario N2J 2Y5, Canada}
\begin{document}
\clearpage \maketitle 
\pagenumbering{roman}
\begin{abstract}

In this paper, we show that the Hamiltonian approach to loop quantum gravity has a fermion doubling problem.
To obtain this result, we couple loop quantum gravity to a free massless scalar and a chiral fermion field, gauge fixing the many fingered time gauge invariance by interpreting the scalar field as a physical clock.  We expand around a quantum gravity state based on a regular lattice and consider the limit where the bare cosmological constant is large but the fermonic excitations have energies low in Planck units. We then make the case for identifying the energy spectrum in this approximation with that of a model of lattice fermion theory which is known to double. 


\end{abstract}
\newpage
\tableofcontents

\pagenumbering{arabic}
\pagestyle{fancy}
\rhead{}
\lhead{\rightmark}

\section{Introduction} \label{intro}

Theories with a discrete notion of space have been a central theme of quantum gravity research since its birth.  Whether assumed, as in Regge calculus or causal set models, used just as a regulator to be removed, as in causal dynamical triangulations, or derived,  as in loop quantum gravity,  in many different approaches to quantum gravity space or spacetime are represented by a discrete structure.  However,  there is a crucial  problem with quantum field theories represented as degrees of freedom on discrete lattices, which is that it is impossible to represent chiral fermions by a local Hamiltonian.  By a chiral fermion we mean one where the representation its left handed fermions live in is not the  complex conjugate of the representation within which the right handed fermions transform. When we attempt to do so, we find that the spectrum of fermonic excitations doubles, with the right handed spectrum being now the mirror of the left handed one.

We know this from examples, which are supported by a powerful no-go theorem proved by Nielsen and Ninomiya\footnote{For the interested reader we review the proof of this theorem in an appendix.} for local quantum field theories on a regular lattice \cite{Nielsen1981, Nielsen1981B, NNProof}.  The result of this theorem is that for such models to be consistent, they must couple equally to the left and right handed sectors of the theory.   As an immediate consequence, the number of left and right handed particles must be matched, and this result is referred to as fermion doubling. This theorem makes the construction of a lattice theory for the standard model quite difficult, since the weak interactions only couple to the left-handed sector of the the theory; if a lattice theory is to be compatible with the well tested experimental evidence of the standard model, it must avoid fermion doubling. 

To date, most research into fermion doubling has been focused on standard quantum field theories. In this paper, we propose that fermion doubling is not unique to quantum field theories, rather we will provide a demonstration of fermion doubling in loop quantum gravity.

One might expect that fermions double in loop quantum gravity just because space is discrete, but there is a deeper reason having to do with the absence of chiral anomalies.  In a now classic series of papers \cite{ThiemannAnomaly96, ThiemannAnomaly98, ThiemannAnomaly982, ThiemannAnomaly983}, Thiemann constructed a regularized form of the Hamiltonian constraint for quantum gravity-either the pure case or the case in which arbitrary collections of Yang-Mills fields, scalars and fermions are included.  In all cases, Thiemann shows that the algebra of quantum constraints is first class and finite, hence, anomaly free. But consider the case in which a chiral gauge field is coupled to a multiplet of chiral fermions which have the feature that, when considered just as a chiral gauge theory, the dynamics would be inconsistent due to a non-vanishing chiral gauge anomaly.  How can coupling to gravity render that inconsistent quantum theory finite and consistent?  The only way this is possible is if fermions double, cancelling the chiral anomaly.

More specifically, we work in the Hamiltonian formulation of loop quantum gravity, coupled to a scalar field and a chiral, left handed fermion field. The chiral fermion may also be coupled to a Yang-Mills gauge field. We follow \cite{RovelliSmolin93, SquareRoot10} and gauge fix the refoliation gauge symmetry by picking the scalar field as a time variable; this gives us a Hamiltonian, ${\cal H}^{QG}$.  We expand the action of that Hamiltonian around a background state of the gravitational field which under coarse graining corresponds to a flat spatial metric.  That background state is constructed from a superposition of spin network states with support on a regular lattice, $\Gamma$.  We then define a Born-Oppenhiemer expansion around that state in addition to an expansion in the inverse of the bare cosmological constant.  (These kinds of expansions were studied previously in \cite{BornOppenheimer1, BornOppenheimer2, BornOppenheimer3} and \cite{SquareRoot10, SquareRoot15}.) We then show that to leading order, the low energy spectrum of ${\cal H}^{QG}$ (for $E \ll E_{Planck}$) is equal to the spectrum of a lattice fermion theory on $\Gamma$.  But we can use the Nielsen-Ninomiya theorem to show that the spectrum of that theory is doubled.  Therefore, the fermonic excitations of the loop quantum gravity state also double.

In a bit more detail, the basic strategy of our analysis is the following:

\begin{enumerate}

\item{} Start with a lattice fermion theory based on a graph $\Gamma$ embedded in $\Sigma$, a torus $T^d$.  Assume $\Gamma$ is regular enough for a fourier expansion of the fermion excitations to exist, allowing there to be a continuous  spectrum $E=E ( p ) $. Verify the assumptions of the Nielsen-Ninomiya no-go theorem hold for the Hamiltonian $H^{LT}_\Gamma$. \label{1}

\item{} Define $\mathcal{H}^{QG}_{\Gamma}$ to be the (spatially diffeomorphism invariant, with a physical hamiltonian coming from gauge fixing) subspace of the Hilbert space of $LQG$ based on a graph $\Gamma$. \label{2}

\item{} Define $H^{\text{matter}}$ to be the Hamiltonian based on graph $\Gamma$ that dominates in the Born-Oppenheimer approximation.  \label{3}

\item{} Use a degravitating map and dressing map to show that the spectra of $H^{\text{matter}}$ and  $H^{LT}_\Gamma$ must be identical, at least in the regime of $E \ll E_{Planck}$. \label{4}

\item{} This establishes the no-go theorem for $LQG$.

\end{enumerate}

The moral of this strategy is to carry the problem of doubling in $LQG$ to a standard lattice fermion theory, where the no-go theorem may be applied. Item 2 grants us a background graph to work on. We use the Born-Oppenheimer approximation to concentrate on the fermionic degrees of freedom and expand around the gravitational pieces, we need the fermions to live on a background graph $\Gamma$ to investigate doubling. Note our result is that there exists $\Gamma$ such that fermion excitations of the corresponding background states double their spectra.  We don't yet show that the fermion excitations around all states double, that would be a stronger result.
 
\section{Fermion Doubling in Loop Quantum Gravity} \label{Proof}

In this section, we will work out the nontrivial \cref{2,3,4} of the program introduced in \cref{intro}.
The main trick is to perform a Born-Oppenheimer like expansion \cite{BornOppenheimer} around a background state which is a functional of the  gravitational degrees of freedom.  We choose to work with a state with support on the spatial diffeomorphism class of a single fixed graph $\Gamma$, which we take to be a regular lattice (we do not expect our results to change if working with a state with support over a countable collection of graphs).  By doing so we map the spectrum of low lying fermonic excitations of the quantum theory to that of a regular fermion theory on a fixed graph, which is the same $\Gamma$. We thus show the spectrum of this theory is equivalent to that of a lattice fermion theory which doubles. To demonstrate this equivalence, we will define degravitating and dressing maps, which carry the Hilbert spaces and Hamiltonians of the two theories into each other. In the low energy limit, the spectra of the two theories become equal.

To reduce quantum gravity to lattice fermion theory, we shall work in the Hamiltonian framework of $LQG$, and study a model where quantum gravity is coupled to a scalar "clock" field $T$ with conjugate momenta $\pi$, plus chiral fermion fields.

We also extend the results to include a lattice Yang-Mills theory with compact gauge group $G$, in which case the fermions are extended to a multiplet of fermion fields in a representation $r$ of $G$ \cite{Rovelli94}. 

The fermion fields are represented by two component spinors $\psi^{A \alpha}$ with conjugate momenta $\pi_{A \alpha}$, where $A = 0,1$ labels components of a Weyl spinor and $\alpha$, which may sometimes be suppressed, labels the basis of the representation $r_0$. For this theory, the Hamiltonian constraint, Diffeomorphism constraint, and Gauss constraint are a sum of terms 
\begin{flushleft}
\begin{align}
C &= C_{\text{grav}} + C_{T} + C_{\psi} +C_{YM}, \,\,\,\,\, D_{a} = D^{\text{grav}}_{a} + D^{T}_{a} + D^{\psi}_{a} + D^{YM}_{a} , \,\,\,\, G_{AB} = G^{\text{grav}}_{AB} + G^{\psi}_{AB}, \nonumber \\
&\text{with} \nonumber \\
C_{\psi} &= \pi^{\alpha}_{A} E^{aA}_{\,\,B} (D_{a} \psi)^{B}_{\alpha}, \,\,\,\,\,\,\,\,\,\,\,\,\,\,\,\,\,\,\,\,\,\,\,\,\,\,\,\,\, D^{\psi}_{a} = \pi^{\alpha}_{A} (D_{a} \psi)^{A}_{\alpha}, \,\,\,\,\,\,\,\,\,\,\,\,\,\,\,\,\,\,\,\,\,\,\,\,\,\,\,\,\,\,\,\,\,\,\,\,\,\,\,\,\,\, G^{\psi}_{AB} = \pi_{(A} \psi_{B)}, \label{time} \\
&\text{and} \nonumber \\
C_{T} &= \frac{1}{2} \pi^2 + \frac{1}{2} E^a_i E^{b i}  \partial_a T\partial_b T, \,\,\,\,\,\,\,\,\, D^{T}_{a} = \pi \partial_{a} T,
 \\
C_{YM}& = \frac{1}{2} qg_{ab} ( e^a e^b + b^a b^b   ), \,\,\,\,\,\,\,\,\,\,\,\,\,\, D^{YM}_{a} = \epsilon_{abc} e^b b^c,
\end{align}
\end{flushleft}
where all constraints with the superscript "grav" are the gravitational counterparts from Hamiltonian general relativity \cite{Rovelli}.

There is also the Gauss's law constraint of the Yang-Mills gauge group, $G$.
\f
G= {\cal D}_a e^a + \pi^A \psi_A,
\ff
where $e^a$ and $b^a$ are the densitized electric and magnetic fields of the Yang-Mills field $a_a$ and here and above we have suppressed the indices for the fermion representation of the gauge group, $r$.

To map a quantum gravity theory to a conventional quantum theory, we need to extract a Hamiltonian from the Hamiltonian constraint.  To do this we gauge fix the many fingered time refoliation invariance, before quantization.  This gauge fixing is accomplished by taking the scalar field $T$ to be a physical clock, so the spacial slices have constant
$T(x) = \tau$.  This turns the Hamiltonian constraint into a Schrodinger-like equation \cite{RovelliSmolin93}
\begin{equation}
i \hbar \frac{\partial}{\partial \tau} \Psi = \int_{\Sigma} d^{3} x \sqrt{- 2[ C_{\text{grav}} + C_{\psi}]} \Psi. \label{FunMatt}
\end{equation}

We will define $\mathcal{W}$ to be the regulated limit of the right-hand side

\begin{equation}
\mathcal{W} = \lim_{L, A, \delta,\tau \rightarrow 0} \int_{\Sigma} d^{3} x \sqrt{- 2 [C^{L A \delta}_{\text{grav}} + C^{L \tau \delta}_{\psi}]}. \label{W}
\end{equation}

The main work we have to do is to define $\mathcal{W}$ acting on spatial diffeomorphism invariant states.  These are states which are functionals of diffeomorphism equivalence classes, which are labelled by the diffeomorphism class of the graph $\Gamma$, which is denoted by $\{  \Gamma \}$.  The state carries additional labels associated with the edges and nodes of $\Gamma$.  
Such states are of the form,
\f
\Psi [\Gamma ] = \braket{\Gamma | \Psi} = \Psi [ \{  \Gamma \} ]
\ff
The regulated operators involved in $\mathcal{W}$ are not diffeomorphism invariant, and so for finite values of the regularization parameters, the action has to be defined on a larger, kinematical Hilbert space,  ${\cal H}^{kin}$.  But, as discussed in detail in 
\cite{SmolinReview}, the limit as the regulators are removed of the operators at issue give us an action purely on ${\cal H}^{diffeo}$.  

We employ a state dependent, graph preserving regularization, so that the action of $\mathcal{W}$ on a state supported on a single graph $\Gamma$ is another state supported on $\Gamma$.  When acting on a state with support on a diffeomorphism equivalence class of a graph
$\{ \Gamma \}$, the action returns a state with support on that diffeomorphism equivalence class \cite{GraphPreserving}.

\subsection{Classical Born-Oppenheimer Approximation}

In this section, we will expand the square root in the operator (\ref{W}). To do this, we separate out the cosmological constant term
\f
 \hat{\cal C}_{\text{grav}} = - \text{det}(E)   \Lambda +  \hat{\cal C}_{\text{Einst}},
\ff
and perform an expansion around large $\Lambda$. Initially, this may look like a very poor expansion. $\Lambda$ is the infrared cutoff  \cite{Smolin95, BMS} of the theory, and expansion around a large $\Lambda$ suggests we are taking the infrared cutoff to be larger than the ultraviolet cutoff. However, this is an expansion around the bare cosmological constant, which we already know to be large by the standard naturalness arguments. In addition, the particular value of $\Lambda$ takes no role in our proof (see \cref{Lattice Born}), and we can add or subtract an arbitrary constant from it at will.

The expansion of $\mathcal{W}$ for large $\Lambda$ is now (with $\epsilon$ as set of regularization parameters \cite{RovelliSmolin93, Rovelli94})
\begin{align}
 \hat{\cal W} &=  \lim_{\epsilon \rightarrow 0}
\int_\Sigma  d^3x \sqrt{\Lambda \text{det}(E) - 2 [ \hat{\cal C}_{\text{Einst}}^\epsilon (x)
+ \hat{\cal C}_{\psi}^\epsilon (x)  ] }  
\\
&= \lim_{\epsilon \rightarrow 0}
\int_\Sigma  d^3x  \sqrt{ \Lambda \text{det}(E) 
\left( 1 - \frac{2}{\Lambda}  \frac{\hat{\cal C}_{\text{Einst}}^\epsilon (x)}{\text{det}(E)}
- \frac{2}{\Lambda} \frac{\hat{\cal C}_{\psi}^\epsilon (x)}{\text{det}(E)}  \right) }
\\
&= \lim_{\epsilon \rightarrow 0}
\int_\Sigma  d^3x  \sqrt{ \Lambda}\left [ \sqrt{   \text{det}(E)}
 - \frac{1}{\Lambda}  \frac{\hat{\cal C}_{\text{Einst}}^\epsilon (x)}{\text{det}(E)^{\frac{1}{2}}}
 - \frac{1}{\Lambda}  \frac{\hat{\cal C}_{\psi}^\epsilon (x)}{\text{det}(E)^{\frac{1}{2}}} + O\left(\frac{1}{\Lambda^2}\right)
\right ]
\\
&= \sqrt{ \Lambda} V -  {\mathcal{W}}^{eff}, \label{W}
\end{align}
where the effective Hamiltonian ${\mathcal{W}}^{eff}$ is
\f
 {\cal W}^{eff}=  \lim_{\epsilon \rightarrow 0} \frac{1}{\sqrt{\Lambda}}\int_\Sigma  d^3x 
 \frac{1}{ \sqrt{   \text{det}(E)}}  \left (
 \hat{\cal C}_{\text{Einst}}^\epsilon (x)   +  \hat{\cal C}_{\psi}^\epsilon (x)
\right )  + O\left(\frac{1}{\Lambda^{3/2}}\right).
\ff

We can formulate the extension of the Nielsen-Ninomiya no-go theorem to LQG by establishing a map between the Hilbert spaces of two theories: $LQG\Psi$, loop quantum gravity with chiral fermions (and possible Yang-Mills fields), and the lattice theory for fermions without gravity.

\subsection{The Quantum Gravity Hilbert Space}

In this section, we shall review some key properties about the quantization. For simplicity, we shall fix the manifold $\Sigma$ to be the torus $T^3$. The Hilbert space of loop quantum gravity coupled to fermions and possible Yang-Mills fields will be denoted
\f
{\mathcal{H}}^{QG\Psi}.
\ff
This is the spatially diffeomorphism invariant physical Hilbert space based on gauge fixing to a constant clock field $T$. On this Hilbert space, we will impose the spatial diffeomorphism and $G$ constraints.  ${\mathcal{H}}^{QG\Psi}$ has a physical and spatially diffeomorphism invariant inner product.

${\mathcal{H}}^{QG\Psi}$ is decomposable in terms of diffeomorphism invariant classes of embeddings of a graph $\Gamma$ into $\Sigma$
\f
{\mathcal{H}}^{QG\Psi}= \oplus_\Gamma {\mathcal{H}}^{QG\Psi}_\Gamma.
\ff
Each component ${\cal H}^{QG\Psi}_\Gamma$ consists of extended spin network states, with labels corresponding to gravity, fermions, and Yang-Mills fields, diffeomorphic to $\Gamma$. One basis to the Hilbert space is the span of all spin-network states \cite{RovelliSmolin95} of the form
\f
\ket{j, i, r, c,\psi},
\ff
where $j$ is the spin of an irreducible representation of the gravitational $SU(2)$ associated to each edge of $\Gamma$, $i$ is an $SU(2)$ intertwiner associated to each node of $\Gamma$, while  $r$ and $c$ are representations and intertwiners of the Yang-Mills gauge group $G$ associated with edges and nodes respectively.
Each node of the graph carries a basis state of the Hilbert space for fermions at a site, carrying a finite dimensional representation of $SU(2) \times G$, spanned by the $\pi$ and $\psi$ operators.

(Technical point, we impose equivalence under point wise smooth diffeomorphisms, so that there are no invariant diffeomorphism invariants characterizing classes of intersections with valence $5$ or greater.)

On ${\mathcal{H}}^{QG\Psi}$, there is a physical Hamiltonian corresponding to the classical operator ${\mathcal{W}}$ (given by \cref{W}), obtained by gauge fixing the clock field to the $T=$ constant gauge. We must now choose a regularization scheme. For simplicity we shall consider a {\it graph preserving regularization}, where ${\cal W}$
decomposes into separate actions on each graph sector
\f
\mathcal{W} = \sum_{\Gamma} \mathcal{W}_\Gamma,
\ff
where each $\mathcal{W}_\Gamma$
acts only on the corresponding $H^{QG\Psi}_{\Gamma}$. We expect the results to be independent of this choice. We shall talk about this regularization choice in \cref{reg}

Also defined on each ${\cal H}^{QG\Psi}_\Gamma$ are algebras of observables, ${\cal O}^{QG\Psi}_\Gamma$, these are defined by loop and flux operators for $SU(2)$ and $G$,  together with the fermion operators $X, Y$, etc, restricted to $\Gamma$.

Given a choice of graph $\Gamma$, there is also a lattice gauge theory involving the same fermion and Yang-Mills fields. The basis states are obtained by simply removing the gravitational degrees of freedom,
\begin{equation}
\ket{r, c, \psi}.
\end{equation} 
The lattice gauge theory Hilbert space ${\cal H}^{GT}_\Gamma$, contains an algebra of operators, ${\cal O}^{GT}_\Gamma$, an inner product, and a Hamiltonian $H^{GT}_\Gamma$.

To summarize, the gravitational, gauge and fermion fields are all represented by a decorated spin network $\Gamma$, whose edged are labelled by $(j,r)$, representations of $SU(2) \times G$, and whose nodes are labelled by intertwiners coming in one of the two types. In the next section, we will discuss the mapping between these two theories.

\subsection{Mapping Quantum Gravity to Lattice Gauge Theory}

In this section, we shall define maps between the two theories for each $\Gamma$, with the goal being to identify the spectrum of $LQG$ based on a graph $\Gamma$ with a lattice gauge theory. The
simplest map to define is the degravitating map
\f
{\cal G}_\Gamma :   \mathcal H^{QG}_\Gamma \rightarrow \mathcal H^{GT}_\Gamma,
\ff
which simply removes the labels corresponding to the gravitational fields
\f
{\cal G}_\Gamma \circ \ket{j,i, r, c,\psi} = \ket{r, c,\psi}.
\ff

To go in the other direction, we define dressing maps
\f
{\cal F}_\Gamma : {\cal H}^{GT}_\Gamma  \rightarrow  {\cal H}^{QG}_\Gamma, 
\ff
which takes a state of the lattice gauge theory and dresses it with gravitational fields; this is defined by a choice of amplitudes $\phi_\Gamma  [ j,i; r,c,\psi ] $ so that 
\f
\mathcal{F}_\Gamma \circ \ket{r, c,\psi} = \sum_{j,i}  \phi_\Gamma [ j,i; r,c,\psi ]  \ket{j,i, r, c,\psi}.
\ff
The relevant choice of these amplitudes will be clear in \cref{Lattice Born}.

We require that 
\f
{\cal G}_\Gamma \circ {\cal F}_\Gamma = I
\ff
for the degravitaing map to be the "inverse" of the dressing map (${\cal G}_\Gamma$ is onto, whereas $ {\cal F}_\Gamma$ is into a subspace).

\subsection{Regulating the Hamiltonian} \label{reg}

We need to define regularizations of the different terms in the Hamiltonian on the fixed graph subspaces $\mathcal{H}^{QG}_\Gamma $ (these are different than those previously mentioned since we are regulating on a fixed graph).  

Let us start with the fermion term in the time-gauge fixed Hamiltonian \cref{time} 
\f
H^\Psi = \int_\Sigma \frac{1}{e} {\cal C}^{\psi} = \int_\Sigma  \frac{1}{e} \, \tilde{\Pi}_{A}^\alpha \tilde{E}^{a A}_{\ \ B} ( {\cal D}_a \Psi )^B_\alpha.
\ff
Acting on a state $\Phi_\Gamma$ in ${\cal H}^{QG}_\Gamma $, we want to regulate this as 
\f
H^\Psi \rightarrow H^\Psi_{reg \ \Gamma} 
= \frac{1}{l_{Pl}}  \sum_{ (n,\hat{a})} 
\Pi (n)_{A}^\alpha \,{E}[S(n,\hat{a})]^{ A}_{\ \ B} U(n,\hat{a})_{B}^{\ \ C}  V(n,\hat{a})_{\alpha}^{\ \ \beta} \,\Psi (n+\hat{a})_{C \beta} ,
\label{gpr}
\ff
with $n$ being the set of nodes and $\hat{a}$ being the corresponding adjacent vertices to each of these nodes.  Both are summed over.
It is the specification of these nodes and edges that make the regularization graph preserving.

Here $E(S)_{AB} $ is the $E^a_{AB}$ smeared over a two surface $S$ with base point $p$ defined by
\f
E(S)_{AB} = 
\int_S d^2 S(\sigma^1, \sigma^2 )_a \,\tilde{E}[S(\sigma )]^a_{CD} \,
U(\gamma_{p , S(\sigma )})_A^{\ C} U(\gamma^{-1}_{p , S(\sigma )})_B^{\ D},
\ff
where $\gamma_{p , S(\sigma )}$ is an arbitrary, non intersecting curve in $S$ connecting the
base point $p$ with the point $S(\sigma )$.
We pick the surfaces $S(n,\hat{a})$ adopted to the graph $\Gamma$ so that for each node $n$ and
adjacent edge $( n,\hat{a}) $, $S(n,\hat{a})$ is a surface that crosses that edge once infinitesimally close to the node $n$ base pointed at the intersection point. 
In addition, $U(n,\hat{a})_{B}^{\ \ C}$ is the $SU(2)$ parallel transport over the edge $(n,\hat{a}) $,
and $V(n,\hat{a})_{\alpha}^{\ \ \beta} $ is the same for the Yang-Mills gauge group $G$. The specific action of the curves and surfaces are crucial to the definition of the regularization.

We also have to define the fermion momentum operator.  Let ${\cal H}^\Psi_n$ be the finite dimensional Hilbert space of fermions at the node $n$, which has a momentum operator $\pi^A_n$, we have,
when acting on a state in ${\cal H}^{QG}_\Gamma $,
\f
\int_\Sigma d^3x\, s^A (x) \tilde{\Pi}_a (x) = \sum_n  s^A (n)  \pi^A_n.
\ff

The graph preserving regularization of the Yang-Mills part of the Hamiltonian works the same way.  We do not give the details here but the important result is that the magnetic field squared term translates to a sum over minimal loops of the product of parallel transports around those loops.  A minimal loop is a c lose path on $\Gamma$ that starts from and arrives to a base point, that is not made up of products of smaller loops.  These are triangles in a simplical lattice and plaquettes in a cubic lattice.  

\subsection{Lattice based Born-Oppenheimer Approximation} \label{Lattice Born}

We now apply the tools discussed earlier in this chapter to identify loop quantum gravity with lattice gauge theory. As mentioned in the outline, we need to perform a Born-Oppenheimer approximation to expand around the gravitational degrees of freedom and work on one particular graph. To do this, we need to define the appropriate background state. Since $\Gamma$ is fixed, we start by defining a class of states which have support only on $\Gamma$
\begin{equation}
\Psi_{\Gamma}[\Delta, i, j] = \braket{0, \Gamma|\Delta, i, j} = \delta_{\Gamma \Delta} f(i, j).
\end{equation}
We must define the background state $\Psi_{0}[\Gamma]$ in this form. We must now choose the particular dependence of the state on the spins $j, i$. The dependence on the intertwiners $i$ and $j$ is chosen so that the state is an eigenvector of the Hamiltonian constraint with the smallest positive energy
\f
 \lim_{\epsilon \rightarrow 0} \frac{\hat{\cal C}_{\text{grav} }^\epsilon (x)}{\text{det}(E)^{\frac{1}{2}}} \ket{0,\Gamma} = E_{\text{min}} \ket{0, \Gamma}. \label{MinGrav}
\ff
In addition, to replicate translation invariance, we will quantize the volume operator
\f
 \lim_{\epsilon \rightarrow 0} \braket{0,\Gamma | \int_\Sigma \sqrt{\text{det}(E) } |0,\Gamma} =v.
\ff
The wave function which satisfies these properties will be referred to as $\Psi_{0}[\Gamma]$. As is done in atomic physics \cite{BornOppenheimer}, we use this choice of background state to perform the expansion

\f
\Psi [\Gamma ]=  e^{\imath \frac{\sqrt{\Lambda}}{c}v} \Psi_0 [\Gamma ] \chi [\Gamma ],
\ff
where the eigenvalue problem for $\Psi_{0}$ fixes $\Gamma, i, j$, $c$ is the coupling constant of the theory, and $\chi$ is a wave function over the fermionic degrees of freedom $\ket{r, c, \psi}$. We point out a subtle issue with the cosmological constant here: we may alter $\Lambda$ and $v$ in such a way as to preserve the following result. The value of $v$ is of no physical importance to us, so instead of taking the IR cutoff to be larger than the UV cutoff, we may add appropriate powers of the UV cutoff to the cosmological constant without distrubing any of our results. To first order in the Born-Oppenheimer approximation, we neglect $\frac{\partial \Psi_{0}[\Gamma]}{d \tau}$, and the resulting Schrodinger equation on a background state is
\begin{align}
{\imath \hbar \partial \chi [ \{ \Gamma  \} , \tau]
\over \partial \tau} &= H^{\text{matter}}  \chi [ \{ \Gamma  \} , \tau]  \\
&\text{with} \nonumber \\ 
H^{\text{matter}} &=  \frac{1}{c \sqrt{\Lambda}}E_{\text{min}} + \frac{1}{c\sqrt{\Lambda}} \lim_{\epsilon \rightarrow 0}  \braket{0,\Gamma |
\int_\Sigma  d^3x \,  \frac{1}{\sqrt{\text{det}(E) } }\,\hat{\cal C}_{\psi, ren }^\epsilon (x)| 0,\Gamma}.
\end{align}

The constant $E_{\text{min}}$ may be subtracted without changing the results. At this stage, we perform the regularization in \cref{reg}.  We evaluate the regulated operator on diffeomorphism invariant states and take the limit as the regularization parameters are removed.  As discussed in \cite{GraphPreserving}, the action is finite on these states and is within the Hilbert space associated with the graph $\Gamma$.   The resulting $LGT$ Hamiltonian is
\f
H^{\Psi GT}_{ \Gamma}   =   \frac{1}{l_{Pl}} 
\sum_{(n,\hat{a})} 
\pi (n)_{A}^\alpha  w^a_{n,\hat{a}} \sigma_a^{AB}  V(n,\hat{a})_{\alpha}^{\ \ \beta} \Psi (n+\hat{a})_{B \beta},
\ff 
where $w^a_{n,\hat{a}}$ is tangent to the edge $(n,\hat{a})$ at $n$ and unit in the background metric,
\f
q^0_{ab} = \sigma_a^{AB} \sigma_{b \ AB},
\ff
as can be seen from the action of the $Y^{a}$ operator on a fixed background. We have now demonstrated one direction of the equivalence of the $LQG$ and $LGT$ theories, namely the degravitating one. We see the lattice spacing, $a$, comes out to be the Planck length.  

The same method applied to the regularization of the Yang-Mills contribution to the Hamiltonian constraint yields the lattice Yang-Mills action, with the lattice spacing coming out as the Planck length.

We must now show the other direction, the dressing map, gives the same spectrum. Once this is complete, we will have shown these two theories are equivalent, hence $LQG$ doubles.

To perform the mapping, we need to define, for each graph subspace  ${\cal H}^{QG}_\Gamma $, the appropriate degrees of freedom for the dressing map
\f
{\cal F}_\Gamma^0 : \chi \in {\cal H}^{GT}_\Gamma  \rightarrow  \chi \times \Phi_\Gamma^0 \in   {\cal H}^{QG}_\Gamma .
\ff
We insist the dressing state $\Phi_\Gamma^0 $ satisfies
\begin{align}
\braket{\Phi_\Gamma^0 |  {E}[S(n,\hat{a})]^{A B}  |\Phi_\Gamma^0} &= w^a_{n,\hat{a}} \sigma_a^{AB} \\
\text{and} \nonumber \\
\braket{\Phi_\Gamma^0 | U(n,\hat{a})_{B}^{\ \ C}  |\Phi_\Gamma^0} &= \delta_B^{\ \ C},
\end{align}

In addition, we also require
\f
\braket{\Phi_\Gamma^0 | \frac{1}{e}  |\Phi_\Gamma^0}=   \frac{1}{e^0} = 1
\ff
to ensure the volumes of cells are equal, as before. Notice we are setting a tensor density with weight one equal to a constant, thus this will only work in some frames.

We note that the $SU(2)$ Gauss's law constraint is not  satisfied.

Once we have defined the dressing and degravitating maps with the above choice of amplitudes, it is easy to show the spectra between the Born-Oppenheimer expanded $LQG$ and the corresponding Lattice Gauge theory are identical
\f
\braket{ \chi |\braket{ \Phi_\Gamma^0 |    H^\Psi_{reg \ \Gamma}    |\Phi_\Gamma^0 }  |  \chi }
=  \braket{ \chi|    H^{\Psi GT}_{ \Gamma}     |  \chi },
\ff
with the lattice spacing taken to be $l_{Pl}$.

This concludes the proof. In summary, we have performed a number of approximations to transform a theory of loop quantum gravity on a background graph into a lattice gauge theory. Upon reading this proof, one may worry about issues such as backreaction. However, it is critical to notice fermion doubling is a low energy phenomena. The left and right handed "limits" of a particle only make sense when the lattice spacing is taken to zero, in $LQG$ the case is identical as the Planck length approaches zero. What makes the proof in $LQG$ a nontrivial exercise is the notion of a background graph $\Gamma$: this is a purely gravitational phenomena and disappears when the Planck length is taken to zero. We must use the background graph for the question of fermion doubling to take any meaning. In addition, the inner product is not the most clearly defined object here, and the concept of momentum space is rather vague. These two points justify our expansions around a background graph and hint towards a mapping to a lattice gauge theory where we know how fermion doubling arises.

\section{Outlook}

In this paper, we have shown the spectrum of loop quantum gravity coupled to free fermions and Yang-Mills fields doubles. This provides an obvious contradiction with well known particle physics, as the weak interactions only couple to the left handed sector of the standard model. In order to connect loop quantum gravity to well-established observations, we need to modify some of the assumptions put into this work. 

The main assumptions can be summarized as follows:
\begin{enumerate}


\item Assume the bare cosmological constant $\Lambda$ is large enough so that the square root in the Hamiltonian constraint can be expanded. 

\item The Born-Oppenheimer approximation is a reasonable assumption: The variation of gravitational fields does not effect the statement of fermion doubling.

\item Assume the interactions between the fermions are local and Hermitian.

\item Assume the momentum space is compact.  

\item Assert chiral invariance has the same definition for both lattice based theories and continuum theories, namely $\{D, \gamma_{5} \} = 0$, with $D$ the Dirac operator.

\item Assume the action is quadratic in the fermionic fields.
\end{enumerate}

We should emphasize that our result does not apply to spin foam models, which are path integral quantizations of loop quantum gravity.  The reason is that the local Lorentz group is exactly realized and is represented by infinite dimensional representations.  Hence, the space of fermionic excitations is non-compact.

Nor does our result rule out the possibility that there may be other backgrounds states of the quantum gravitational field, expansions around which lead to a chiral fermion spectrum that does  not double.  Gambini and Pullin have presented an example in the case of a $1+1$ dimensional reduction of loop quantum gravity  \,\cite{gp}.  We note that their example involves a continuous superposition of background lattices, and will not apply to any superpositions of a countable number of (diffeomorphism invariant classes) of graphs.

This highlights a missing element from loop quantum gravity research, which is the absence of a Hamiltonian for the gravitational sector, which is bounded from below \cite{PEQG}.  Without this, we can construct semiclassical states which course grain to flat space, but we have no method to determine which is the best approximation to the ground state.

Let us then consider how we might modify the Hamiltonian theory in a way that avoids fermion doubling. We believe the first four assumptions are reasonable. In particular, fermion doubling is a low energy artifact of the theory, where it should be safe enough to ignore Planck-scale physics, and the cosmological constant can be added and subtracted at will. Instead, we suspect the problem is due to the choices related to the matter sector. Below, we will discuss several possible strategies to avoid fermion doubling. 

One of the requirements for the Nielsen-Ninomiya theorem to hold is that the fermion Hamiltonian must be quadratic in the fields. In standard $QFT$, this is a perfectly reasonable assumption to make as higher order terms are irrelevant; in 4D the coefficients of a higher order term in the fields has a positive mass dimension, hence they're irrelevant. However, in a lattice theory there is a natural cutoff provided, and in loop quantum gravity UV divergences are completely avoided, so in principle such terms can be added to the action at will. One possibility is that there may be a higher order action which does not double and is suitable for loop quantization. 

Another possibility to cure the problem would be to adapt known techniques from lattice gauge theory, such as the Ginsparg-Wilson technique, among others (these include Domain-Wall Fermions, Overlap Fermions, Twisted Mass QCD, Staggered Fermions, Wilson Fermions, etc, see~\cite{GinspargWilson, DomainWallFermions, Overlap, TwistedMassQCD, Stagger, Creutz, WilsonFermions}). We could use the dressing map to define a Born-Oppenheimer expanded theory of quantum gravity which does not double, and use this to motivate a chiral regulation to the Hamiltonian constraint \cref{FunMatt}.

A third, and related, possibility is to look at perfect actions \cite{PerfectAction}. In this framework a lattice theory can exactly replicate the full set of gauge symmetries the continuum limit of the theory possesses. This is done through a process of coarse-graining the fields. However, this comes at the price of locality. In addition, results can currently only be obtained in a perturbative framework, which is seemingly against the logic of $LQG$.

Finally, we may hypothesize that chiral fermions are emergent states of the quantized gravitational field, such as the chiral braided states proposed in \cite{braid}.  In these states the chirality of the excitations is a consequence of the chirality of the braids themselves.
\section*{ACKNOWLEDGEMENTS}

We are grateful to Rodolfo Gambini  and Jorge Pullin for correspondence on these results.
This research was supported in part by Perimeter Institute for Theoretical Physics. Research at Perimeter Institute is supported by the Government of Canada through Industry Canada and by the Province of Ontario through the Ministry of Research and Innovation. This research was also partly supported by grants from NSERC, FQXi and the John Templeton Foundation.

\appendix

\section{Appendix:  Review of fermion doubling in lattice models} \label{FD}
\subsection{Review of Chiral Symmetry}

In this appendix we review the proof of the Nielsen-Ninomiya no-go theorem presented in \cite{Nielsen1981B}. Other, more rigorous proofs are listed \cite{Nielsen1981, NNProof}. Before getting carried away with details, we shall take a small amount of time to discuss what fermion doubling physically means.

In Quantum Field Theory courses \cite{PeskinSchroeder}, we are taught massless solutions to the dirac equation are eigenstates of the helicity operator
\begin{equation}
\hat{h} = \hat{p} \cdot \vec{S} = \hat{p}_{i} \left(\begin{array}{cc} \sigma^{i} & 0 \\ 0 & \sigma^{i} 
\end{array}\right),
\end{equation}
with $\sigma_{i}$ being the Pauli Matrices. A particle with helicity $h =$ +/- $\frac{1}{2}$ is referred to as a right/left handed particle. In other words, a right-handed particle's momentum vector is aligned with its spin (and anti-aligned for a left-handed particle). While this is perfectly consistent and clear in the normal continuum setting, in the discrete setting a particle must come with both helicities. To see this, we look at the fourier kernel $e^{ipx/\hbar}$ to notice a minimum length corresponds to a maximum momentum. Thus the momenta $p$ and $p + \frac{2 \pi a}{\hbar}$ are identified, with $a$ being the lattice spacing. We may add subtract off as many units of $\frac{2 \pi a }{\hbar}$ as we like, resulting in a positive momenta vector being identified with a negative one. Under this identification, the helicity changes sign and we see left and right handed particles must be equivalent.

Another way to arrive at fermion doubling is to look at the dispersion relation in momentum space. Since momentum space is bounded, any functions on it must be periodic. Thus the dispersion relation will cross the line of zero energy up and down an equal number of times. 
On each occasion, the continuum dispersion relation $E=pc$ emerges in the low energy limit; however, it will come with varying sign and thus varying helicity. Since there are an equal number of zero energy crossings with positive and negative slopes, there are an equal number of emergent left and right handed fermions. As a result, fermion doubling occurs in all lattice theories with compact (closed and bounded) momentum space. This is the key feature we will look at in our proof.

Helicity is not a meaningful observable if a particle is massive. This can be summarized quite simply, a massive particle does not travel at the speed of light, and hence we can consider an observer which overtakes the particle. In this frame, the direction of the momentum is flipped, and the helicity changes sign. For massive particles, it is convenient to think of a more abstract concept, known as chirality. The chirality of a particle is determined from whether the particle transforms under the left or right-handed part of the Poincare group. From here, we may work with chirally invariant actions, where the action has a symmetry under
\begin{equation}
\psi_{L} = e^{i \theta_{L}} \psi_{L}, \,\,\,\, \psi_{R} = e^{i \theta_{R}} \psi_{R},
\end{equation}
where we can define the left and right handed pieces using the projectors $P_{L} = \frac{1-\gamma_{5}}{2}$, $P_{R} = \frac{1+\gamma_{5}}{2}$. While chirality is a reasonably defined concept for massive particles, it is easy to show mass terms break chiral symmetry. To do this, we can write a mass term using the left and right-handed parts
\begin{equation}
m \bar{\psi} \psi = m (\bar{\psi}_L \psi_L + \bar{\psi}_L \psi_R + \bar{\psi}_R \psi_L + \bar{\psi}_R \psi_R).
\end{equation}
We see the cross terms break chiral invariance, and a chirally invariant theory must be massless.
For a massless theory, chirality is conserved for a massless particle and is equal to its helicity.

\subsection{Nielsen Ninomiya no-go Theorem}
We are now prepared to formulate the no-go theorem:
\begin{theorem}
Suppose we are given a lattice theory of free fermions governed by an action quadratic in the fields
\begin{equation}
S = -i \int dt \, \sum_{x} \dot{\bar{\psi}}(x, t) \psi(x, t) - \int dt \, \sum_{x, y} \bar{\psi}(y, t) H(x-y) \psi(x, t),
\end{equation}
with $\psi(x, t)$ an N component spinor with discrete position label x.
Then fermion doubling is inevitable if the following conditions are met: \\
$\bullet$ The underlying lattice has a well defined momentum space which is compact. \\
$\bullet$ The interaction $H$ is hermitian and local, in the sense that its momentum space continuation is continuous. \\
$\bullet$ The charges $Q$ are conserved, locally defined as a sum of charge densities $Q = \sum_{x} j^{0}(x) = \sum_{x} \bar{\psi}(x) \psi(x)$, and is quantized. \\
\end{theorem}
The proof below can be generalized to a more general kinetic term \, $\dot{\bar{\psi}}(y, t)T(x-y) \psi(x, t)$ so long as the term $T$ doesn't vanish anywhere on the lattice, as can be seen from the equations of motion.
\begin{equation}
i \sum_{y} T(x-y) \partial_{t} \psi (y, t) = \sum_{y} H(x-y) \psi(y). \label{EOM}
\end{equation}
Inverting the $T$ matrix, the equations of motion indicate the action is equivalent to one governed by the standard kinetic term along with the effective Hamiltonian $H/T$; if $T$ is singular or nonlocal the no-go theorem does not apply. Because of this it is often convenient to speak in terms of the Dirac operator $D$, defined by
\begin{equation}
S = a^{4} \sum_{x} \bar{\psi} D \psi,
\end{equation}
where now we add an additional assumption to the theorem: $D$ is required to be invertible.

In addition, we may couple to the fermions a fixed gauge field (such as a Yang-Mills field) without spoiling the proof; such fields simply act as a background when working in the low-energy regime.

To characterize fermion doubling, we will look at the dispersion relation in momentum space, determined from the eigenvalue problem
\begin{equation}
H(p) \psi_{i}(p) = \omega_{i}(p) \psi_{i}(p),
\end{equation}
here the index $i$ is not to be confused with the components of $\psi$, but rather a label of eigenvectors which is not summed on. We can choose this label such that the eigenvalues are increasing (recall they are real by the Hermiticity of $H$)
\begin{equation}
\omega_{1}(p) \le \omega_{2}(p) \le \dots \le \omega_{N}(p).
\end{equation}
The relevant feature will be captured by the level crossings
\begin{equation}
\omega_{i}(p_{deg}) = \omega_{i + 1}(p_{deg}),
\end{equation}
as this corresponds to the zeroes of the energy difference in our simple picture. We will now show that the level crossings correspond to left or right handed Weyl particles. 

As an aside, notice that level crossings between 2 levels are generic in 3+1 dimensions, but 3 level crossings are not. To see this, consider a general 3x3 Hamiltonian spanned by the Gell-Mann matrices $\lambda_{i}$
\begin{equation}
H^{(3)}(p) = A(p) \mathbb{1} + \sum_{i = 1}^{8} B_{i}(p) \lambda_{i}.
\end{equation}
For a three level degeneracy the coefficients $B_{i}(p)$ must all vanish. This requirement gives 8 equations for only 3 quantities, which generically has no solutions. However, two level crossings are generic in 3+1 dimensions as the relevant 2x2 piece of the Hamiltonian is spanned by three Pauli matrices. 

Near a 2-level crossing at the degeneracy point $p_{\deg}$, we may expand the relevant piece of the Hamiltonian in a Taylor Series
\begin{equation}
H^{(2)}(p_{\text{deg}} + \delta p) = \omega_{i}(p_{\text{deg}}) \mathbb{1} + \delta \vec{p} \cdot \vec{A} + \delta p_{k} V^{k}_{\alpha} \sigma^{\alpha} + O(\delta p^2)
\end{equation} 
where $\alpha$ and k run from 1 to 3. To simplify the form of this Hamiltonian near $p_{\deg}$, we will shift the momentum by
\begin{align}
P_{0} = H &\rightarrow P_{0} - \omega_{i}(p_{\text{deg}}) - \delta \vec{p} \cdot \vec{A} \\
\delta p_{k} &\rightarrow \delta p_{k} \pm \delta p_{\alpha} V^{\alpha}_{k}, 
\end{align}
where the $\pm$ sign depends on the determinant of $V$. The shifted Hamiltonian becomes
\begin{equation}
H = \vec{p} \cdot \vec{\sigma},
\end{equation}
and the corresponding eigenvalue problem is
\begin{equation}
\hat{\vec{p}} \cdot \vec{\sigma} U(p) = \pm p_{0} U(p), \label{Weyl}
\end{equation}
where $U(p)$ is the corresponding wave function. The sign of the determinant of $V$ is thus identified with the helicity of a Weyl particle. Thus each 2 level-crossing $p_{\text{deg}}$ is identified with a Weyl fermion in the low energy limit. It is important to note the helicity depends on the degeneracy point $p_{\text{deg}}$ and so it now remains to identify which types of Weyl fermions emerge in the low energy limit.

To set about this task we will look at curves in the 3-D dispersion relation space (which is embedded in the full 3+1D $\omega$-$p$ space) defined by
\begin{equation}
\{(p, \omega)| \braket{a|\omega_{i}(p)} = 0 \},
\end{equation}
where the bracket is
\begin{equation}
\braket{a|\omega_{i}(p)} = a_{1} \psi_{1}^{(i)} +  a_{2} \psi_{2}^{(i)} + \dots +  a_{N} \psi_{N}^{(i)},
\end{equation}
with $\ket{a}$ being any constant N-vector. In 3 dimensions, this will turn into 1 complex equation for 3 quantities $\vec{p}$, implying the set forms a curve. These curves are of special importance because they pass through all of the degeneracy points $p_{\text{deg}}$. The reason for this is that at a degeneracy point, we may redefine the energy eigenstates through any superposition of the form
\begin{equation}
\ket{\omega_{i}} = \alpha \ket{\omega_{i}} + \beta \ket{\omega_{i+1}}.
\end{equation} 
Thus we may choose $\alpha$ and $\beta$ so that $\braket{a|\omega_{i}(p_{\text{deg}})} = 0$. All such curves must be closed since the Brillouin zone is compact. Thus if the curves have an orientation, they must pass equally many times up and down through the degeneracy points, proving the theorem. In the paragraphs below we will set up this orientation.

To set about this task, we will look at the phase of $\braket{a|\omega_{i}}$ on small circles wrapping around the curve near each degeneracy point. Suppose we set this small circle of radius $R$ a distance $d$ away from a degeneracy point at $p_z = 0$ along a curve passing through the positive $p_z$ direction. The Weyl equation near the degeneracy point is given by \cref{Weyl}, with the sign being identified with helicity. The eigenvectors to this equation along the mentioned circle $S^{1}  = \{\theta \in \mathbb{R} | (p_x, p_y, p_z) = (R \cos{\theta}, R \sin{\theta}, d)\}$ in the limit $R \ll d$ are
\begin{equation}
U_{1} = \left( \begin{array}{c}
1 \\ \frac{R}{2 d} e^{i \theta}
\end{array} \right), \, \omega_{1} = \pm (d + R), \, U_{2} = \left( \begin{array}{c}
1 \\ -\frac{R}{2 d} e^{-i \theta}
\end{array} \right), \, \omega_{2} = \mp (d + R).
\end{equation}
In the limit $R \rightarrow 0$, we have $U_{1, 2} = \left( \begin{array}{c}
1 \\ 0
\end{array} \right)$, implying the vector $\bra{a}$ along the curve is $(0, 1)$. We obtain the following for the phase rotation along the curves for both levels
\begin{equation}
\braket{a|U_{1}} = \pm \frac{R}{2 p_z} e^{i \theta}, \, \braket{a|U_{2}} = \mp \frac{R}{2 p_z} e^{- i \theta} \label{orient}.
\end{equation}
\cref{orient} may now be used for orientation assignment. There are then two types of curves for each helicity, one crossing from $p_0 < 0, p_z < 0$ to $p_0 > 0, p_z > 0$, and one crossing from $p_0 < 0, p_z > 0$ to $p_0 > 0, p_z < 0$. Using \cref{orient}, we see the first curve is oriented in the positive (negative) $p_z$ direction for a right (left) handed degeneracy point and the second is oriented in the negative (positive) $p_z$ direction for a right (left) handed degeneracy points. Since the curve must be closed and passes equally many times up and down through degeneracy points, by the orientation assignment we see the number of left and right handed particles are matched. \qed

\clearpage
\pagestyle{empty}
\renewcommand{\thepage}{}
\bibliography{mybib}
\end{document}